\documentstyle[twocolumn,prl,aps,epsfig,
]{revtex}

\catcode`@=11
\def\citer{\@ifnextchar [{\@tempswatrue\@citexr}{\@tempswafalse\@citexr[]}}
 
\def\@citexr[#1]#2{\if@filesw\immediate\write\@auxout{\string\citation{#2}}\fi
  \def\@citea{}\@cite{\@for\@citeb:=#2\do
    {\@citea\def\@citea{--\penalty\@m}\@ifundefined
       {b@\@citeb}{{\bf ?}\@warning
       {Citation `\@citeb' on page \thepage \space undefined}}%
\hbox{\csname b@\@citeb\endcsname}}}{#1}}
\catcode`@=12

\def\beq{\begin{equation}}
\def\eeq{\end{equation}}

\def\beqn{\begin{eqnarray}}
\def\eeqn{\end{eqnarray}}
\relax
\newcommand{\ba}[1]{\begin{array}{#1}}
\def\ea{\end{array}}

\def\be{\begin{equation}}
\def\ee{\end{equation}}
\def\bea{\begin{eqnarray}}
\def\eea{\end{eqnarray}}

\def\to{\rightarrow}

\def\dis{\displaystyle}
\def\f{\frac}
\def\[{\left[}
\def\]{\right]}
\def\({\left(}
\def\){\right)}
\def\nn{{/\hspace*{-1.83mm}}}
\def\ppbar{p\bar p}

\def\sq2{\sqrt{2}}
\def\gaga{\gamma\gamma}

\def\G{{\cal G}}
\def\ee{e^+e^-}
\def\ff{f\bar{f}}
\def\qq{q\bar{q}}
\def\llbar{\ell^-\ell^+}
\def\vv{\nu\bar{\nu}}
\def\tt{t\bar{t}}
\def\MM{M_{\star}}
\def\pbinv{pb$^{-1}$}
\def\fbinv{fb$^{-1}$}
\def\lumi{\int\!{\cal L}}
\def\End{\end{document}}
\def\thisday{\today ~and~ hep-ph/9904220~~}
\begin{document}                                                              
\draft

\twocolumn[\hsize\textwidth\columnwidth\hsize\csname
@twocolumnfalse\endcsname
     
\title{   
        Collider Tests of Compact Space Dimensions             
                  Using Weak Gauge Bosons  
}  
\author{\sc Csaba Bal\'azs,~~ Hong-Jian He,~~ 
            Wayne W. Repko,~~ C.--P. Yuan}
%\address{\phantom{ll}}
\address{
Department of Physics and Astronomy,
Michigan State University, East Lansing, Michigan 48824, USA 
}
\author{\sc Duane A. Dicus}
\address{
Center for Particle Physics and Department of Physics, 
University of Texas, Austin, Texas 78712, USA
}
\date{\thisday}
\maketitle
\begin{abstract}
\hspace*{-0.35cm}
We present collider tests of the recent proposal for  
weak-scale quantum gravity due to new large compact space 
dimensions in which only the graviton ($\G$) propagates. 
We show that the existing high precision LEP-I $Z$-pole 
data can impose non-trivial constraints on the scale of the new 
dimensions, via the decay mode $Z\to f\bar{f}+\G$ ($f=q,\ell$). 
These bounds are comparable to those obtained at high energy colliders 
and provide the first sensitive probe of the scalar graviton.
We also study $W(Z)+\G$ production and the anomalous $WW(ZZ)$ 
signal from virtual $\G$-states at the Fermilab Tevatron, and compare 
them with the LEP-I bound and those from LEP-II and future linear colliders. 
\linebreak
{PACS number(s): 04.50.+h, 11.25.Mj, 14.70.-e   \hfill [MSUHEP-90105]}
\end{abstract}
\vskip1pc]

\setcounter{footnote}{0}
\renewcommand{\thefootnote}{\arabic{footnote}}

%\vspace*{0.25cm}
\noindent
%\underline{\it 1. Introduction} 
%\vspace*{0.2cm}

The smallness of the Newton constant,
$G_N\simeq 1/(1.2\times 10^{19}\,{\rm GeV})^2$, 
suggests that the characteristic
scale for the gravitational interaction is the Planck scale 
$M_P=1/{\small \sqrt{G_N}}$. This traditional wisdom holds, however,
only if gravitons ($\G$) effectively propagate in the usual 
$4$-dimensional space-time all the way to 
Planck scale, where string theory provides 
the ultraviolet-finite unification
of gravity with other three gauge forces of the Standard Model (SM).
In $D=4+n$ dimensions, it is possible to introduce a new Planck scale 
$M_{\star}$ and 
compact $n$-dimensional volume $R^n$ related to the usual $4$-dimensional
Planck scale by Gauss' law \cite{model} as
\beq
M_{\star}^{n+2} R^n = M_P^2 / 4\pi \,.
\label{eq:gauss}
\eeq
The new gravitational scale 
$M_{\star}$ can be as low as the weak scale,
e.g., $M_{\star}\!\sim\! 1$\,TeV for $R\!\sim\! 1$\,mm and $n=2$.
This provides an intriguing way to resolve or reinterpret the hierarchy
problem and opens an exciting opportunity for testing quantum
gravity in the TeV regime. 
As recently pointed out in the literature \cite{model}, 
such millimeter-range space dimensions are not in contradiction
with any existing macroscopic measurement of the gravitational force. 
Allowing only the graviton, but not SM fields, 
to propagate into the extra dimensions 
is theoretically natural \cite{witten}.
It is found that, irrespective 
of the detailed underlying dynamics at high
scales (which is currently neither unique nor predictive),
the $4$-dimensional effective theory below the new Planck scale $M_{\star}$ 
is essentially the SM plus additional interactions of its
fields with a tower of massive Kaluza-Klein (KK) excitations of the
graviton coupled to the energy-momentum tensor\,\cite{model,GRW,peskin,HLZ}.
These KK modes, with masses $m_\ell =|\ell |/R$, have tiny
mass-separations $\delta m_\ell \!\sim\! 1/R$ 
(which is $\sim\! 2\times 10^{-4}$\,eV for $R\!\sim\! 1$\,mm).
This means a summation over all possible KK states is necessary for 
computing physical processes. Practically, this summation may be replaced
by a continuous integral over the KK states with a proper density
function. The crucial feature is that the effective 
gravitational interaction strength after KK-summation is $1/\MM$
(instead of $1/M_P$) so that it is testable for 
$\MM = O({\rm TeV})$\,\cite{model}-$\!\!$\cite{new1}
\footnote{In the following, 
unless specified otherwise, the symbol $\G$ always
denotes the graviton plus its KK modes and the summation over KK
is implied. For the spin-0 case, $\G_0$ denotes  the scalar
KK tower.}.

In this paper, we first analyze a direct probe of both spin-0 and 
spin-2 KK excitations via the decay $Z\to \ff+\G$ ($f=q,\ell$) 
and derive non-trivial bounds on the scale $\MM$ (and $R$) 
from the existing high precision $Z$-pole 
data at CERN LEP-I\,\cite{LEPI-data}. 
We then study $W(Z)+\G$ production and the anomalous $WW(ZZ)$ signal
at the Fermilab Tevatron. These results are compared with the LEP-I bound
and those from $\ee\!\to\! Z+\G$ at LEP-II and future Linear Colliders 
(LCs). The unique role of LEP-I for probing  the scalar 
KK modes via $ZZ\G$ coupling and the importance of the Tevatron 
for testing $n\geq 3$ are stressed.

\vspace*{0.15cm}
\noindent
\underline{\it $Z\to\ff+\G$ and High Precision LEP-I $Z$-Pole data} 
\vspace*{0.1cm}

Thus far, most analyses of the direct or indirect tests 
for the existence of $\G$  focus
on production processes 
such as $\ff\to\gamma + {\G},\,{\rm jet}+{\G}$ and 
$\ff \to ({\G}^\ast )\to \ff ,\gaga$, 
etc\,\cite{peskin}-$\!\!$\cite{add}. 
In these processes, only the spin-2 graviton $\G_2$ is relevant 
since the scalar graviton $\G_0$ coupling to
matter fields is proportional to their masses and 
is thus vanishing for massless gauge bosons (such as the photon or gluon)
or negligible for light fermions. 
As shown below, the scalar KK modes of graviton are best probed 
using the $Z$-decay into $\ff +{\G}$ at LEP-I.\footnote{Other 
possible {\it high energy}
processes, such as $\ff\to W(Z)+\G$ and
$VV\to VV,\tt$ ($V=W,Z$) at hadron/lepton colliders, 
and $\ee\to WWZ,ZZZ$ at LCs, may also probe $\G_0$. But 
the contribution of $\G_0$ is suppressed relative to that of $\G_2$
at high energies, as to be addressed later in the text.}
%It is interesting to note that all these processes also 
%need to make use of at least one $VV\G$ coupling.} 
Studying this process probes a different aspect of the
dynamics of weak scale quantum gravity.

In the effective 4-dimensional theory below the scale $\MM$, the graviton
plus KK modes universally couple to SM fields via the
energy-momentum tensor and its trace\,\cite{GRW,peskin,HLZ}, and the effective
Lagrangian
is 
\beq
{\cal L}_{\rm eff}^{\G} = -\f{\kappa}{2}
\left[\omega\G_0T^\mu_\mu+\G_2^{\mu\nu}T_{\mu\nu} \right]\,,
\eeq
where $\kappa = \sqrt{32\pi G_N}$ and
$\omega = 1/\sqrt{3(n+2)/2}$\,\cite{HLZ}.
(We have used the same conventions for $\MM$ and $R$ as in
Ref.\,\cite{peskin}.)
The relevant $T_{\mu\nu}$ tensors for $Z$-fermion interactions are
\beqn
T^{Z}_{\mu\nu} &=& -Z_{\mu}^\alpha Z_{\nu\alpha}+M_Z^2Z_\mu Z_\nu 
                   +\f{g_{\mu\nu}}{4}\[ Z_{\alpha\beta}^2
                   -2M_Z^2Z_\alpha^2 \]\,, \\
T^{f}_{\mu\nu} &=& 
 \f{1}{4}\[\bar{\psi}\gamma_{\mu}iD_{\nu}\psi  
  -(iD_{\nu}^{\dagger}\bar{\psi})\gamma_{\mu}\psi\]
  +(\mu\leftrightarrow\nu )\,,
\eeqn
where $D_{\mu}$ denotes the SM gauge covariant derivative and
$Z_{\alpha\beta}=
\partial_{\alpha}Z_{\beta} - \partial_{\beta}Z_{\alpha}$. 
To include $W$'s, we need only add the corresponding
$T_{\mu\nu}^W$ tensor.

Consider the decay $Z(p)\to f(k_1)+\bar{f}(k_2)+{\G}(p')$.
For the scalar graviton $\G_0$, there is 
only one diagram with $Z \to Z^\ast +\G$ followed by $Z^\ast \to
\ff$. In the case of a spin-2 graviton $\G_2$, there are 
additional graphs with $Z \to \ff$ and $\G_2$ emitted from
either the  $f$ or $\bar f$, as well as a contact graph containing the 
$Z$-$f$-$\bar f$-$\G_2$ vertex. 
The decay amplitudes are
\begin{eqnarray}
%\begin{array}
{\cal A}_{\G_0} &=& \dis i\kappa \bar{u}(k_1)X_\alpha
      (g_V-g_A\gamma_5)v(k_2)\epsilon_Z^\alpha(p),
\\[0.6mm]
{\cal A}_{\G_2} &=& \dis i\kappa\bar{u}(k_1)X_{\alpha\mu\nu}
       (g_V-g_A\gamma_5)v(k_2)
       \epsilon_Z^\alpha(p)\epsilon_{\G}^{\mu\nu}(p'),
\end{eqnarray}
\vspace*{-8mm}
\begin{eqnarray}
X_\alpha &=& 
\dis M_Z^2\omega\gamma_{\alpha} /(s-M_Z^2),
\nonumber \\%[0.5mm]
\dis X_{\alpha\mu\nu} &=& 
\dis\f{1}{s-M_Z^2}\[p_\mu p_\nu\gamma_\alpha\hspace*{-1.25mm}
-\hspace*{-0.3mm}g_{\mu\alpha}p_\nu\nn{p} 
\hspace*{-0.23mm}+\hspace*{-0.5mm}
(p_\mu p'_\alpha \hspace*{-1.3mm}-\hspace*{-0.1mm}g_{\mu\alpha}
p\!\cdot\!p' )
\gamma_\nu \] \nonumber \\%[1mm]
&& \hspace*{-10mm} \dis-\f{1}{2}
\[
\!\f{1}{t}\gamma_\alpha 
(\nn{k}_1\!\!-\!\nn{p})k_{2\nu}\gamma_\mu\!\! +
\f{1}{u}\gamma_\mu k_{1\nu}(\nn{k}_2\!-\!\nn{p})\gamma_\alpha    
 \!\!-\! g_{\nu\alpha}\gamma_\mu\],
\nonumber
\end{eqnarray}
where $s=(k_1+k_2)^2,~t=(p-k_1)^2,~u=(p-k_2)^2$, 
and $\epsilon_Z^\mu (\epsilon_\G^{\mu\nu} )$
is the polarization vector of $Z^\mu \,(\G_2^{\mu\nu})$. In the above,
$\gamma_{\alpha}(g_V-g_A\gamma_5)$ represents the $Z$-$f_i$-${\bar f}_i$ 
SM coupling, i.e.,
%\begin{eqnarray}
$\(g_V,\;g_A\)  =  
\(I_{3i} - 2Q_i\sin^2\theta_W\!,\;I_{3i}\)(g/2\cos\theta_W)\,, 
$
%\end{eqnarray}
with $I_{3i}$ denoting the third component of weak isospin for the $i$th
fermion $f_i$ and $Q_i$ its electric charge.
The partial decay width is given by
\beqn
\Gamma\left(Z\to \ff+\G_i\right)=\dis\f{1}{256\pi^3 M_Z^3}
\int\hspace*{-2.5mm}\int\hspace*{-2.5mm}\int
\overline{|{\cal A}_{\G_i}|^2}\,ds\,dt\,d{\cal N}\,,
\label{eq:widint}
\eeqn
where $d{\cal N}=\rho (m)dm^2$ with $\rho (m )$ 
denoting the KK states density function 
$~\rho (m)= \pi^{n/2}R^n m^{n-2}/\Gamma (\f{n}{2})$ 
\cite{GRW,peskin,HLZ}.~
$\overline{|{\cal A}_{\G_i}|^2}$ is the squared, spin-averaged amplitude
with $\ff$ final states summed for leptons and light quarks
In the case of $\G_0$, we find
\beqn
\overline{|{\cal A}_{\G_0}|^2}=
\dis\f{4}{3}\[\f{g_x\kappa\omega M_Z}{s-M_Z^2}\]^2
\(2p\cdot k_1\,p\cdot k_2+M_Z^2k_1\cdot k_2\)\,,
\eeqn
where $g_x^2=g_V^2+g_A^2$. 
The $d{\cal N}$ integration 
forces the integrand to vanish as $s\to M_Z^2$.
The integral Eq.\,(\ref{eq:widint}) can be evaluated numerically
and the results cast into the following form
\beqn
\(
\matrix{\Gamma\left(Z\to\hspace*{-1mm}\ff\hspace*{-1mm}+\G_0\right)\nonumber\\
        \Gamma\left(Z\to\hspace*{-1mm}\ff\hspace*{-1mm}+\G_2\right)\nonumber}
\hspace*{-10.7mm}\)
\hspace*{-1mm}\f{1}{\,\Gamma_0} =\dis\f{1}{8\pi}
\( 
\matrix{2\omega^2\nonumber\\
        1       \nonumber} 
\hspace*{-10.7mm}\)
\hspace*{-1mm}\(\f{M_Z}{\MM}\)^{\!\!n+2}
\hspace*{-1.5mm}\(
\matrix{I_{n0}\nonumber\\
        I_{n2}\nonumber} 
\hspace*{-10.7mm}\)
\\[2mm]
 =  \(
\matrix{0.80\times 10^{-7}/{\MM}^{\!4}\nonumber\\
        1.66\times 10^{-7}/{\MM}^{\!4}\nonumber} 
\hspace*{-10.7mm}\),~~({\rm for}~n=2)\\ [-5mm]
\label{eq:wid-G02}
\eeqn
where $\MM$ is in TeV and
$\Gamma_0$ 
is the SM decay width of $Z\hspace*{-1mm}\to \hspace*{-1mm}\ff$.
$I_{ni}$ is an integral depending on $n$ 
and the spin of $\G_i$.
In the case of $\G_0$, we have
\beqn
&I_{n0}&=\dis\!\frac{\pi^{(n-2)/2}}{\Gamma(n/2)}\!\!
\int_0^1\hspace*{-2mm}
%\int_0^{(1-\sqrt{x})^2}\!\!
%\f{y^{{(n-2)}/{2}}\,(12x+A)\,\sqrt{A}}{6(1-x)^2}
%\,dxdy
\int_0^{(1-\sqrt{x})^2}\hspace*{-16pt}dxdy\,
\f{y^{{(n-2)}/{2}}\,(12x+A)\,\sqrt{A}}{6(1-x)^2}
\eeqn 
with $A=(1-x-y)^2-4xy$. Eq.\,(\ref{eq:wid-G02})
shows that, for $n=2$, the $\ff+\G_0$ channel contributes 
about $1/3$ of the new partial decay width while $\ff+\G_2$ channel 
about $2/3$. This provides the first effective test of
the coupling of the scalar graviton $\G_0$.
For $n=(2,3,4)$, we find 
\beqn
\dis\f{\Gamma\left(Z\!\to\!\ff\!+\!\G_{0\oplus 2}\right)}{\Gamma_0}\!=\!
\(\f{2.46}{{\MM}^4},\,\f{.075}{\MM^5},\,\f{.0029}{{\MM}^6}
\)\!\times\! 10^{-7}\,,
\eeqn
which decreases rapidly for $n\geq 3$.
This is due to the power suppression of the $(M_Z/\MM)^{n+2}$ factor in 
Eq.\,(\ref{eq:wid-G02}) and can only be improved by going to energies
above the $Z$ pole, as discussed below.

On the other hand, the high precision LEP-I experiments
have accumulated a sample of about $2.3\times 10^7$
on-shell $Z$ boson decays via the $\qq$ and $\llbar$ 
channels \cite{LEPI-data,LEPIx}. 
The SM background for our study is
the rare decay channel $Z\!\to\! \ff\!+\!\vv$. 
In fact, ALEPH performed a
partial analysis for an integrated luminosity $\int\!{\cal L}=79$\,pb$^{-1}$ 
data sample and found no events above the SM prediction\,\cite{LEPI-data}.
To estimate how the whole LEP-I data sample can constrain the graviton
signal, we calculate the SM partial decay width of 
$Z\!\to\! \ff\!+\!\vv $ ($f=q,\ell$) and derive its
decay branching ratio as
\beqn
{\rm BR}[Z\!\to\! \ff\!+\!\vv ]
&=& \!\f{(4.269_{\qq\vv}+0.779_{\llbar\vv})
\!\times\! 10^{-7}\,{\rm GeV}}
{2.494_{\rm total}\,{\rm GeV}}  \nonumber
\\
&\simeq& 2.02 \times 10^{-7}. 
\eeqn
We expect about
$(2.3\times 10^7)\times (2.0 \times 10^{-7})\simeq 4.6$ 
background events from the SM. Assuming only 5 events show up in 
$\ff +$missing channel from the whole LEP-I data sample, we deduce,
according to Poisson statistics\,\cite{poisson}, there are about $6$ 
signal events allowed at $95\%$C.L. in the  $\ff+\G$ channel. 
The $95\%$C.L. LEP-I bounds on $\MM$ can be obtained from
\begin{equation}
\(\f{2.46}{{\MM}^4},\,\f{.075}{\MM^5},\,\f{.0029}{{\MM}^6}
\)\!\times\! 10^{-7} \leq  \f{6}{2.3\times 10^7\times 0.8}\,,
\end{equation}
yielding, for $n=2,3,4$, 
\begin{equation}
\MM  \geq  932,\;470,\;310\;{\rm GeV}\,,
\label{eq:Zbound}
\end{equation}
where the SM branching ratio 
BR$[Z\!\to\! \ff\,]\simeq 0.8$ (for $f=q,\ell$) is used.
From Eq.\,(\ref{eq:gauss}), the above bound 
implies $\,R\leq 0.77$\,mm at $95\%$C.L.
We further note that the SM $\ff+\vv$ events have a very
different topology from the $\ff+\G$ signal, as shown in Fig.\,1
for the energy distribution $dN/d(E_f+E_{\bar f})$ with 5 background
events and 6 signal events. 
Using these distinct signal/background distributions, one can
further improve the bound and possibly push $\MM$ above $1$\,TeV, if a signal 
is not found.
We see that the existing LEP-I high precision $Z$-pole data can
already put non-trivial {\it direct} constraint on $\MM$ (or $R$) 
that is comparable to other bounds obtained
for various high energy colliders  \cite{GRW}-$\!\!$\cite{new1}. Based on these
encouraging results, we conclude that it is important to
extend the existing ALEPH analysis\,\cite{LEPI-data} 
to the total LEP-I $Z$-pole data sample 
for this channel. This should
provide a sensitive direct probe of $\MM$.
\vspace*{-1.3cm}       
\begin{figure}[H]
%\hspace*{-1.0cm}
\epsfig{file=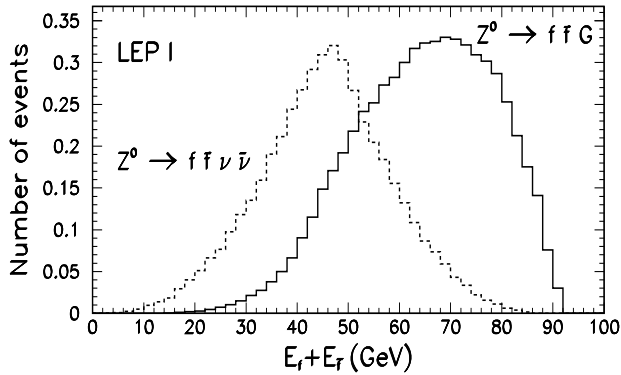,width=8.5cm,height=6cm}
\vspace*{2mm}
\caption[fig:fig1]{
Energy distribution $dN/d(E_f+E_{\bar f})$ of
the signal $\ff+\G$ and background $\ff+\vv$ 
in the $Z$-decay at LEP-I.
}
\label{fig:fig1}
\end{figure}
\vspace*{-0.35cm}

The above analysis can be extended to the case
of on-shell $Z\!+\!\G$ production at LEP-II. However, as 
shown in Ref.\,\cite{new2}, for $\G=\G_2$ and $n=2$, 
the current data can only set a $95\%$C.L. bound on for this mode 
(in analogy to the $ZH$ search mode with invisible Higgs decay)
of $\MM\geq$\,364 GeV,   
where $\MM$ has been converted into the same definition as 
Ref.\,\cite{peskin} and ours.
It is found that
to obtain a bound on $\MM$ of order $\agt 1$\,TeV,
using the $\! Z+\!\G$ channel, a machine with 
higher energy and luminosity is needed, which is only feasible
at future LCs with $\sqrt{s}\geq 500$\,GeV and 
$\int\!{\cal L}\geq 50$\,fb$^{-1}$\,\cite{new2}.
Our analysis of $Z\!+\!\G$ production 
at LEP-II and LCs confirms this conclusion.

\vspace*{0.15cm}
\noindent
\underline{\it $W(Z)+\G$ and $WW(ZZ)$ Production at Tevatron} 
\vspace*{0.1cm}

As pointed out above, to extend our LEP-I study and especially
to probe $\MM$ for the number of new dimensions larger than 2,
it is desirable to go significantly beyond the $Z$-pole energy.
Before the operation of the Large Hadron Collider (LHC),
the upgraded Tevatron will have the highest collider energy and 
luminosity available to probe $\MM$ in the processes
$\ppbar\to W(Z)+\G$ and 
$\ppbar\to q\bar{q}'\to (\G^\ast )\!\to WW\,\mbox{\rm or}\,ZZ$
\footnote{At the LHC, the sensitivity is expected to be much better and
the large gluon fusion $gg\to Z+\G ,WW(ZZ)$ 
should be included as well.}.
The partonic amplitude and cross section of $\qq\to Z+\G$ can be
obtained from our $Z$-decay result by crossing
and the extension to  $\qq '\to W\!+\!\G$ can also be derived.

For high energy collider production with $V$'s
($V=W,Z$), such as $V\G$ or $VV$, 
the naive expectation is that the longitudinal polarization  
$V_L $ may help to enhance the cross section 
due to the leading energy-dependence of the longitudinal polarization vector 
{\small $\epsilon_L^\mu (k) = k^\mu/M_V+O(M_V/E)$}. 
This however turns out not to be the case for 
gravitational $\G$-coupling 
to the SM fields via the energy-momentum tensor $T^{\mu\nu}$
since the conservation of $T^{\mu\nu}$ requires $k_\mu T^{\mu\nu}=0$.
Consequently, the longitudinal contribution is suppressed at high
energies because
{\small $\epsilon_L^\mu (k) - k^\mu/M_V=O(M_V/E)\ll 1$.}
As a result, when probing the gravity scale $\MM$, only the 
$V_T$-polarizations dominate at high energies.
Nevertheless, the momentum-dependent non-renormalizable coupling of
$\G_2$ still leads to 
direct production cross sections with a 
large energy-enhancement factor, behaving as
$\sim\!(\sqrt{s})^n/\MM^{n+2}$. Hence, a large energy is crucial for 
compensating
the $1/\MM $ power suppression (especially for larger $n$).
The coupling of $\G_0$ to $V$ is 
proportional to $M_{V}$ so that its contribution
is not enhanced when $M_{V}/\sqrt{s}\ll 1$.

We first consider the direct production of $W(Z)+\G$ at the Tevatron
($\sqrt{s}=1.8$\,TeV and $\int\!{\cal L}=100$\,\pbinv) 
and its upgrade ($\sqrt{s}=2$\,TeV and $\int\!{\cal L}=2$\,\fbinv). 
The dominant SM backgrounds are $\qq '\to W(Z)+\vv$,
in which $\vv$ mainly comes from $Z$-decay. 
Since the SM rates of $WW/ZZ/WZ$-pairs are not large at the Tevatron, 
these backgrounds are not serious. We consider only the leptonic decay modes 
of $W(Z)$.
To effectively suppress the fake backgrounds from misidentification of jets,
we require that the $W(Z)$ bosons have transverse momentum $P_T\geq 25$\,GeV 
and rapidity $|y|\leq 2$.
With these cuts, we find the SM $W(Z)+\vv$ background cross sections,
without including the branching ratios, to be about
0.35(0.32) and 0.51(0.38)\,pb   
at the 1.8 and 2.0\,TeV Tevatron(TEV).  The signal
$W(Z)+\G$ cross sections (in fb) as a function of $\MM$ and for 
$n=(2,4,6)$ are 
\beqn
{\rm TEV(1.8)}\!:&& 
166(145)/\MM^4,\,59(62)/\MM^6,\,24(28)/\MM^8;\\
{\rm TEV(2.0)}\!:&& 
241(212)/\MM^4,\,108(112)/\MM^6,\,54(65)/\MM^8\,.
\label{eq:TeV-Sigma}
\eeqn
To derive our Tevatron bounds, we use an
estimated systematic error of $\!\sim\!\! 10\%$ 
in the cross section measurement.
The $95\%$C.L. bounds 
on $\MM$\,(in TeV) for $n=(2,4,6)$ are found to be
\beqn
{\rm TEV(1.8)}\!:~~&&  \MM  ~\geq~  .89(.76),\, .78(.72),\, .67(.71),\\
{\rm TEV(2.0)}\!:~~&&  \MM  ~\geq~  1.2(1.1),\, .98(.98),\, .90(.92).
\label{eq:VGTEV-bound}
\eeqn
For comparison, we have also studied $\ee\to Z+\G$ at
the LC\,(0.5\,TeV,   {\small $\int\!{\cal L}$}$=50$fb$^{-1}$)
and LC\,(1\,TeV,   {\small $\int\!{\cal L}$}$=200$fb$^{-1}$)
with the angular cut 
{\small $|\cos\theta_Z|\leq 0.8$} and 
invariant mass cut 
{\small $M_{\G}=(s-2E_Z\sqrt{s}+M_Z^2)^{1/2}\geq 200$}\,GeV
\cite{new2}. 
(Our numerical results for the $Z+\G$ cross 
sections at the LCs agree with Ref.\,\cite{new2} 
after taking into account the difference in conventions.) 
To derive the LC bounds, we include both the hadronic and leptonic 
($ee,\mu\mu$) decay modes of $Z$, and assume
an identification probability of $74\%$ for $Z$ 
via dijet mass reconstruction\,\cite{Barger} 
as well as  a $2\%$ systematic error for the cross section
measurement\,\cite{LEPI-data}. The SM $Z+\nu\bar{\nu}$ cross section is 
found to be
203\,fb and 512\,fb for the LC(0.5 TeV) and LC(1.0 TeV).
The $95\%$C.L. bounds on $\MM$\,(in TeV) for $n=(2,4,6)$ are
\beqn
{\rm LC(0.5)}\!: ~~&&  \MM  ~\geq~  1.9,\,\, 1.3,\,\, .99,\\
{\rm LC(1.0)}\!: ~~&&  \MM  ~\geq~  2.8,\,\, 2.2,\,\, 1.8.
\label{eq:VGLC-bound}
\eeqn
With a 90\% right-hand polarized electron beam, the SM background rate is 
reduced by a 
factor of 10 due to the suppression of the $W$-$W$ fusion contribution, while 
the 
graviton signal is reduced by only about 20\%. This leads to new $95\%$C.L. 
bounds on $\MM$\,
(in TeV) for $n=(2,4,6)$ of
\beqn
{\rm LC(0.5)}\!: ~~&&  \MM  ~\geq~  2.7,\,\, 1.6,\,\, 1.2,\\
{\rm LC(1.0)}\!: ~~&&  \MM  ~\geq~  4.4,\,\, 3.0,\,\, 2.3,
\label{eq:VGLCPOL-bound}
\eeqn
which clearly demonstrate the importance of having a polarized electron beam.

$WW(ZZ)$ pair production via virtual $\G_2^\ast$ exchange
at the Tevatron can also probe the
new Planck scale $\MM$. 
The leading contribution comes from the 
interference of $s$-channel $\G_2^\ast$-exchange with the
SM terms. For $\MM\gg\sqrt{s}$, the $\MM$-dependence of the interference term 
is {\small $\MM^{-4}\ln(\MM/\sqrt{s})$} for $n=2$ and 
{\small $\MM^{-4}$} for $n\geq 3$\,\cite{HLZ},
and the effect tends to decrease the SM rate. 
In this study, we only consider local operator effects and take the logarithm in
the $n=2$ case to be 1 \cite{GRW,add}.
To derive bounds on $\MM$ via $WW/ZZ$ production at the Tevatron, we consider 
both the 
lepton plus jet and the di-lepton modes of $WW$ pairs as well as the pure 
leptonic 
decay modes of $ZZ$ pairs.
Cuts of $P_T\geq 20$\,GeV 
and $|y|\leq 2$ are imposed on each $W(Z)$.
With these cuts, the SM cross
sections for $WW(ZZ)$ production are about 7.1(0.88)pb and 8.1(1.0)pb at
1.8 and 2.0\,TeV, respectively.
The contributions to the cross sections (in fb) from the $1/\MM^4$ terms for
 $n=(2,4,6)$ are
\beqn
{\rm TEV(1.8)}\!:&&   
\, -(180(66),\,280(100),\,220(81))/\MM^4,\\
{\rm TEV(2.0)}\!:&&   
\, -(230(86),\,370(130),\,290(110))/\MM^4.
\label{eq:Sigma-VV}
\eeqn
The 95\%\,C.L. bounds on $\MM$ (in TeV) for $n = (2,4,6)$ are
\beqn
{\rm TEV(1.8)}\!:~~&&   \MM  ~\geq~  .57(.39),\, .64(.44),\, .61(.42),\\
{\rm TEV(2.0)}\!:~~&&   \MM  ~\geq~  .73(.59),\, .82(.66),\, .77(.63).
\label{eq:VVTEV-bound}
\eeqn
For $WW$-production at the 2 TeV Tevatron, the assumed systematic error 
dominates the  statistical 
error, so that studying the di-lepton modes of $WW$ pairs alone gives about 
the same bounds on $\MM$.
These bounds can be compared with 
those from $\ee\to WW(ZZ)$ at
    LEP-II\,(0.2TeV, $\lumi =2$fb$^{-1}$)
and LCs.
With the acceptance cuts $|\cos\theta_V |\leq 0.9$ 
for $V=W$ or $Z$, the SM cross sections of $WW(ZZ)$
pair production at LEP-II(0.2), LC(0.5) and LC(1.0) are about
$16(1.2)$, $2.5(0.19)$ and $0.59(0.043)$\,pb. 
Again, the effect of $\G_2^\ast$ contribution is to 
decrease the SM rate.
Our cross section results agree with those of Ref.\,\cite{new1}, 
although our bounds differ owing to the choice of luminosities and our 
inclusion of systematic errors, which dominate the statistical errors for 
$WW$-production.
Here, the $95\%$C.L. bounds on $\MM$\,(in TeV) for $n=(2,4,6)$ are
\beqn
{\rm LEP2(0.2)}\!:~~&&  
\MM  ~\geq~  .69(.82),\, .77(.92),\,.73(.86),\\
{\rm LC(0.5)}\!: ~~&&   
\MM  ~\geq~  1.7(2.3),\, 1.9(2.6),\, 1.8(2.4),\\
{\rm LC(1.0)}\!: ~~&&   
\MM  ~\geq~  3.4(4.6),\, 3.8(5.1),\, 3.6(4.8).
\label{eq:VVLC-bound}
\eeqn

In conclusion, collider tests of weak-scale quantum gravity using weak gauge
bosons can provide important bounds on the scale parameter $\MM$. 
We showed that LEP-I data can already impose useful constraints on $\MM$ 
which are
comparable to those obtained from higher energy collider studies. Searches at   
the upgraded Tevatron and future LCs using single/double 
$W(Z)$-production can further push the bounds above 
$1$\,TeV, if a signal is not found. Such searches also allow us to probe 
new dimensions with $n\geq 3$, which is difficult at LEP-I.

\smallskip 
%\noindent
We thank M. Kobel for discussing the $Z$-pole
data at LEP-I\,\cite{LEPI-data,LEPIx}.
This work is supported in part by the National Science Foundation 
under grants PHY-9802564 and PHY-9802439 and 
by the U. S. Department of Energy under Contract No. DE-FG013-93ER40757. 
\\[-9mm]

%\end{narrowtext}
\end{document}